\documentclass[aps,prl,twocolumn,showpacs,groupedaddress,amsmath,amssymb]{revtex4}

\usepackage{graphicx}
\usepackage{color}
\usepackage{comment}
\usepackage{bm}
\usepackage{latexsym}

\usepackage[symbol]{footmisc}
\usepackage{hyperref}
 \usepackage{braket}

\usepackage[T1]{fontenc}
\usepackage[latin9]{inputenc}
\usepackage{amsmath}
\usepackage{esint}

\begin{document}

\title{Statistical Design of Chaotic Waveforms with Enhanced Targeting Capabilities}
\author{Huanan Li$^{*}$, Suwun Suwunnarat\footnote{These two authors contributed equaly. HL performed the theoretical analysis 
and SS the numerical analysis}, Tsampikos Kottos}
\address{Physics Department, Wesleyan University, Middletown CT-06459, USA}

\date{\today}

\begin{abstract}
We develop a statistical theory of waveform shaping of incident waves that aim to efficiently deliver energy at weakly lossy targets 
which are embedded inside chaotic enclosures. Our approach utilizes the universal features of chaotic scattering -- thus minimizing 
the use of information related to the exact characteristics of the chaotic enclosure. The proposed theory applies equally well to 
systems with and without time-reversal symmetry.  
\end{abstract}

\pacs{Valid PACS appear here}

\maketitle

The prospect of utilizing waveform shaping of incident acoustic or electromagnetic radiation in order to efficiently 
direct energy to focal points, placed inside chaotic (or disordered) enclosures, has been intensely pursued during 
the last few years \cite{MLLF12}. The excitement for this line of research is twofold: On the fundamental side the 
interesting question is to identify schemes that will allow us to utilize multiple scattering events in complex media 
like disordered structures or chaotic reverberation cavities in order to overcome the diffraction limit \cite{BPZ01,
FR02,VLM10,PABVLM11,KSBS11,CYYKLDFC11,HLFL16,HLGCS17}. A successful outcome can revolutionize 
many applications of wave focusing in complex media, including imaging techniques \cite{FCDPRTTW00,BPTB02,
MRDNF02,CPCLD04} medical therapies \cite{HRY15}, outdoor or indoor wireless communications \cite{MBBSS00} 
and electromagnetic warfare \cite{DRJF10}. 

In this endeavor, the time reversal (TR) and wave front shaping (WFS) are among the most promising wave-focusing 
schemes with impressive experimental demonstrations in a range of frequencies (see review \cite{MLLF12}). 
Disregarding subtle details, it was shown that the wave focusing process is benefited from the multiple scattering 
events and reverberation occurring during propagation inside a complex medium \cite{MLLF12,FR02,ABBKKR17,
RG17}. These two methods are complementary in the sense that TR is a broadband approach which results in 
spatiotemporal focusing of incident waves while WFS is mainly a monochromatic concept which results in maxima 
of deposited energy at desired foci. Both methods, however, are requiring a time-reversal invariance of the propagation 
medium--a condition that can be violated either because of inherent losses or because of some external magnetic 
field. Moreover, they are not addressing the fact that in typical circumstances, where the targets are inside complicated 
enclosures, the scattering fields demonstrate an extreme sensitivity to the exact configuration of the enclosure, 
its coupling to the interrogating antennas, the operating frequency etc. Thus, in many practical applications, the 
design of a waveform with 100\% focusing efficiency (for a {\it specific} configuration of an adiabatically varying 
enclosure) is unfortunately irrelevant. 

At the same time, there are well developed statistical methods, applicable both in the frame of electrodynamics 
\cite{HJ,GYXAAO14} and acoustics \cite{TS07,WW10,LS91}, whose central theme is the statistical description of 
wave interferences at any position inside a complex enclosure \cite{ABF10,S99}. The guiding viewpoint of this school 
of thought is that the (adiabatic) changes of the enclosure render any attempt to describe transport characteristics 
for a specific "replica" of the system meaningless. Thus, the statistical description constitutes the only meaningful 
approach for the scattering properties of chaotic enclosures. Obviously this approach does not provide any recipe 
for the realization of incident waveforms that will lead to foci (hot-spots) inside a complicated enclosure. 

In this paper we develope a {\it statistical} approach for the design of Waveforms with Enchanced TArgetted Capabilities 
(WETACs), that have high probability to deliver large amount of their carrying energy at weakly lossy targets which are 
placed inside complex enclosures. Our scheme distributes the injected energy carried by a WETAC over multiple channels 
and utilizes the features of chaotic scattering as they are quantified by the so-called Ericson parameter. The design of 
WETACs requires a {\it minimal} information about the enclosure: the loss-strength (conductivity) of the target (with some 
tolerance); the eigenfrequencies and the eigemmode amplitudes of the isolated cavity at the positions of the interrogating 
antennas and at the target(s). While most of these informations are experimentally accessible via reflection measurements, 
the field amplitude at the position of the target might not be easily measured. The additional assumption that the latter is 
given by the ergodic limit of field intensities for chaotic systems is proven successful, particularly for the case of multiple 
lossy targets. Finally we find that the WETAC algorithms are applicable for systems with and without TR invariance. 

{\it Modeling Absorption in Complex Enclosures --} A chaotic enclosure is typically modeled by an ensemble of $N\times N$ 
random matrices $H_0$ with an appropriate symmetry: in the case that the isolated cavity is TR invariant, $H_0$ is taken 
from a Gaussian Orthogonal Ensemble (GOE), while in the case where the TR-symmetry is violated it is taken from a Gaussian 
Unitary Ensemble (GUE) \cite{ABF10,S99}. When (multiple) localized, weakly lossy target(s) with loss-strength $\gamma_d$ 
are present, the system is described by the Hamiltonian \cite{LSFSK17,FSK17}: 
\begin{equation}
H(\{\gamma_d\})=H_{0}-\imath\Gamma_{0};\quad\Gamma_{0}=\sum_{d}\gamma_d\left|e_{d}\right\rangle \left\langle e_{d}\right|,
\label{eq: close_sys}
\end{equation}
where $\left\{ \left|e_{l}\right\rangle \right\} $ is the basis where $H_{0}$ is written and $d=1,\cdots N_d$ indicates the lossy 
target index. We shall assume below that all targets have the same losses i.e. $\gamma_d=\gamma$.

We turn the isolated cavity to a scattering set-up by coupling it with $M$ semi-infinite single-mode leads. The leads feature
one-dimensional tight-binding dispersion $E(k)=2t_{L}\cos k$ where $k\in [-\pi,\pi]$. The coupling of the cavity with the leads 
is controlled by the $N\times M$ matrix $W$ with elements $W_{lm}=w\delta_{lm}$ ($w$ is the coupling strength). 

The object that characterizes the scattering process of an incident monochromatic wave with energy $E(k)$ (where 
$k\in (0,\pi]$) is the $M\times M$ scattering matrix $S(k,\gamma)$
\begin{equation}
\mathcal{S}\left(k,\gamma\right)=-\hat{1}+2\imath\frac{\sin k}{t_{L}}W^{T}\frac{1}{H_{eff}(k,\gamma)-E(k)}W,
\label{ScatteringM}
\end{equation}
where $\hat{1}$ is the $M\times M$ identity matrix and $H_{eff}$  
\begin{equation}
H_{eff}(k,\gamma)=H(\gamma)+\frac{e^{\imath k}}{t_{L}}WW^{T}
\label{Heff}
\end{equation}
is an effective Hamiltonian that describes the cavity in the presence of radiative and Ohmic losses.

In the absense of any losses, the scattering matrix is unitary i.e. $S^{\dagger}(k,\gamma=0)S(k,\gamma=0)=1$. However 
when $\gamma\ne 0$,  the scattering matrix is sub-unitary and one can define an absorption operator $\mathcal{A}\equiv 1-\mathcal{S}^{\dagger}\mathcal{S}=\mathcal{A}^{\dagger}$ \cite{F03}. Using Eq. (\ref{ScatteringM}) we get (see supplement 
\cite{supplement})
\begin{align}
\mathcal{A}\left(k,\gamma\right)= & -4\gamma\frac{\sin k}{t_{L}}\sum_{d}\left|u_{d}\right\rangle \left\langle u_{d}\right|,
\:\left|u_{d}\right\rangle =W^{T}G\left|e_{d}\right\rangle 
\label{A_matrix}
\end{align}
where $G\left(k,\gamma\right)=\left[H_{eff}^{\dagger}\left(k,\gamma\right)-E\left(k\right)\right]^{-1}$. 

The absorbance associated with an incident waveform $\left|\mathcal{I}\right\rangle$ is defined as
\begin{equation}
\label{absorbance}
\alpha\left(k,\gamma\right)\equiv{\left\langle \mathcal{I}\right|A\left(k,\gamma\right)\left|\mathcal{I}\right\rangle 
\over \left\langle \mathcal{I}\right.\left|\mathcal{I}\right\rangle}\in \left[0,1\right]
\end{equation} 
An $\alpha=1$ indicates that the energy carried by the incident waveform is completely absorbed by the target(s). The opposite 
limit of $\alpha=0$ corresponds to an incident waveform that "lost" completely the lossy target(s) and has been either transmitted or (and) 
reflected by the cavity. Below, we shall use $\alpha$ as a measure of success of a designed waveform to deliver its energy to a lossy target.

\begin{figure}[h]
\includegraphics[width=1\columnwidth,keepaspectratio,clip]{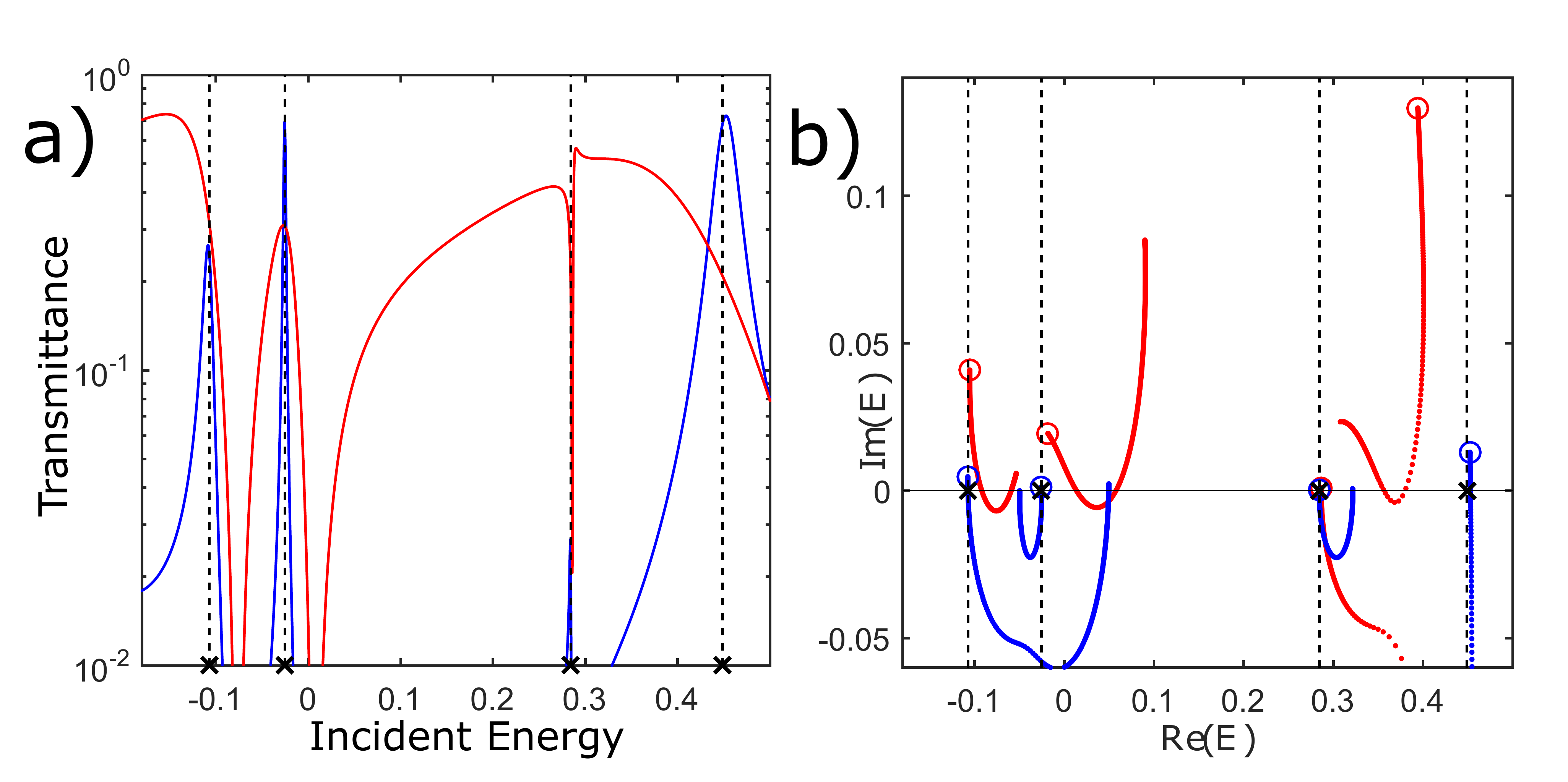}
\caption{ (color online) (a) The transmittance versus energy for a GOE cavity with one loss target of loss-strength $\gamma=0.01$. 
(b) The parametric evolution of complex zeros of the secular equation Eq. (\ref{secular}) in the complex energy plane, as the loss-
strength $\gamma$ increases. In both cases the GOE matrix $H_0$ has dimensionality $N=15$, the number of leads is $M=2$ and 
the coupling constant of the leads is $t_L=-1$. Blue (red) lines indicate a system with a coupling constant between the leads and the 
cavity which is $w = -0.2$ ($w=-1$), with a corresponding Ericson parameter ${\mathcal E} \approx 0.03$ (${\mathcal E} \approx 
1.4$) indicating isolated (overlapping) resonances. The complex zeros for $\gamma=0$ are indicated with open blue/red circles for
each case respectively. Black crosses indicate the position of the eigenenergies $E_n^{(0)}$ of the isolated Hamiltonian $H_0$.}
\label{fig1}
\end{figure}

{\it Perfect Waveforms --}A perfect waveform (PW) corresponds to an incoming wave whose energy is completely absorbed by 
the lossy target(s). The PW $\left|\mathcal{I}_{PW}\right\rangle$ is an eigenvector $\left|\alpha(k_{PW},\gamma)\right\rangle$ 
of ${\mathcal A}(k_{PW},\gamma)$ with a corresponding eigenvalue $\alpha(k_{PW},\gamma)=1$. This condition defines the 
{\it real-valued} wavevector $k_{PW}$. The reality of the wavevector is a physical requirement and it is associated with the fact 
that, in order to transport energy, the input signal has to be a propagating wave. It deserves to point out that PWs have recently 
attracted a lot of attention in the framework of optics where they have been identified as the time-reversed of a lasing mode 
\cite{CGCS10,WCGNSC11}. While these studies are restricted to integrable cavities with TR-symmetry, PWs can also emerge 
in chaotic systems with or without TR-symmetry \cite{LSFSK17,FSK17}.

It is straightforward to show that, for a fixed $\gamma$, $k_{PW}$ are the {\it real} zeros (if exist) of the secular equation $\det 
\left[\mathcal{S}(k,\gamma)\right]=0$. Using Eq. (\ref{ScatteringM}) one can rewrite the secular equation in terms of the effective 
Hamiltonian Eq. (\ref{Heff}) as \cite{LSFSK17,FSK17}
\begin{equation}
\label{secular}
\zeta\left(k,\gamma\right)\equiv\det\left(H_{eff}(-k,\gamma)-E(k)\right)=0.
\end{equation}
Note that, for a fixed $\gamma$, the secular equation $\zeta(k_n,\gamma)=0$ has in general multiple complex zeros $k_n$. 


\begin{figure}[h]
\includegraphics[width=1\columnwidth,keepaspectratio,clip]{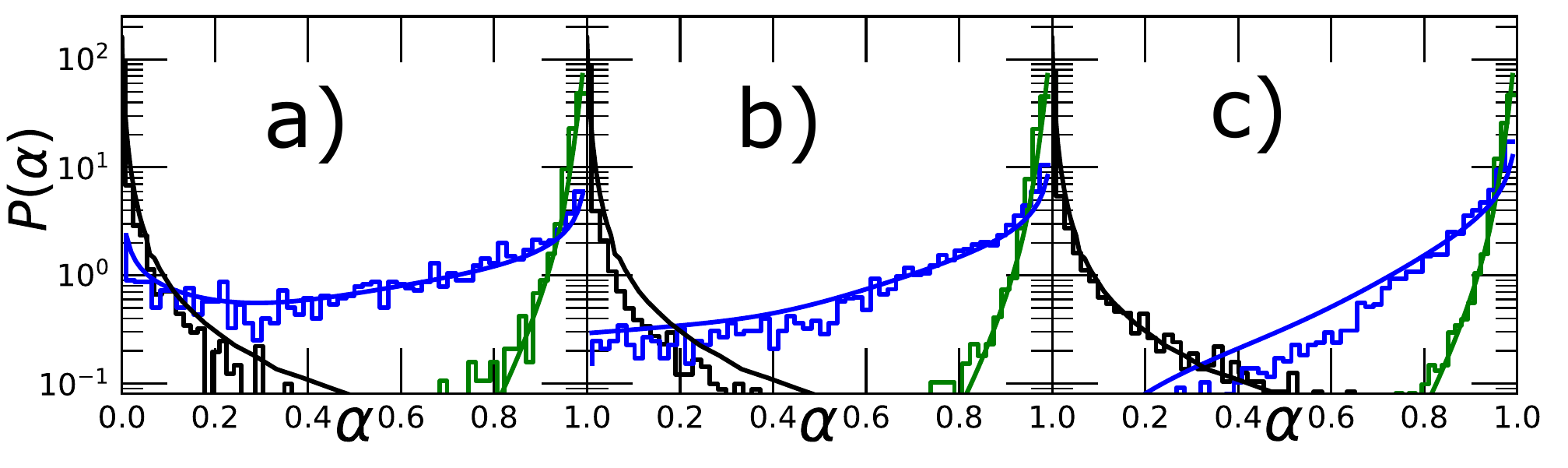}
\caption{ (color online) Numerical (staircase lines) and theoretical (smooth lines) distribution of absorbances ${\mathcal P}(\alpha)$ 
for a chaotic cavity with (a) one; (b) two and; (c) four lossy targets. Blue (green) lines are associated to WETAC (ergodic WETAC) 
incident waveforms. Black lines are associated with an ensemble of $\left|\mathcal{I}_R\right\rangle$ incident waveforms. In all 
cases $M=2, t_L=-1, w=-0.3$ (corresponding to ${\mathcal E}\approx 0.06$). The dimensionality of $H_0$ is $N=15$ while ${\bar 
\gamma} = 0.01$ and $\eta = 0.35$.}
\label{fig2}
\end{figure}

{\it Characterization of chaotic PW based on Ericson Parameter --}The scattering properties of a chaotic cavity depend crucially on 
the way that the system is coupled to the leads. In the case of weak coupling, the scattering matrix exhibit fluctuations on the level 
of the mean level spacing $\Delta$ of the corresponding isolated system \cite{S99}. Furthermore, the transmittance consists of 
resonances that demonstrate narrow linewidths $\Gamma_n$ which is typically smaller than $\Delta$, see blue line in Fig. \ref{fig1}a. 
In the opposite limit of strong coupling, the scattering matrix elements develop universal fluctuations due to interference effects between 
multiple overlapping resonances \cite{S99}. Specifically, the scattering amplitudes and transmittances can be represented by a sum 
of many overlapping resonances, see red line in Fig. \ref{fig1}a.

The distinction between these two qualitative different scattering domains is typically described by the so-called Ericson parameter 
${\mathcal E}$ which is defined as the ratio of the mean resonance width $\langle\Gamma\rangle$ to the mean level spacing 
$\Delta$ of the energy levels of the corresponding isolated cavity i.e. ${\mathcal E}=\langle \Gamma\rangle/\Delta$. When 
${\mathcal E}\ll 1$ the resonances are well isolated from one another while in the opposite case we have strongly overlapping 
resonances.

It turns out that the Ericson parameter controls the nature of the PW as well. In Fig.\ref{fig1}b we show the parametric evolution
of the complex zeros in the ${\mathcal R}e(E)-{\mathcal I}m(E)$-plane as the loss-strength $\gamma$ increases. At the same
figure we also mark with crosses the eigenvalues $\{E_n^{(0)}\}$ of the Hamiltonian $H_0$. Initially (i.e. for $\gamma=0$) the 
zeros are in the upper part of the complex plane (see blue and red circles) because of causality. As $\gamma$ increases they 
move downwards and eventually cross the real axis at $E_{PW}\equiv E(k_{PW})$ corresponding to a critical value of $\gamma
=\gamma_{PW}$. It is exactly this pair of $(E_{PW},\gamma_{PW})$ for which a PW can be achieved. Notice that when ${\mathcal 
E}\ll1$ (blue trajectories), the $E_{PW}$ (whenever they exist) are very close to the eigenvalues $\{E_n^{(0)}\}$ of the isolated 
system. In the opposite limit of ${\mathcal E}\gg 1$, the PW energies $E_{PW}$ occur between two nearby energy levels 
$\{E_n^{(0)}\}$ indicating that more than one mode might affect their formation.

{\it Design schemes for WETACs --}We start our analysis with the observation that a WETAC can be determined by a subset of 
the normalized eigenmodes $\left|\Psi_n^{(0)}\right\rangle$ of the Hamiltonian $H_0$. The size ${\mathcal N}$ of this subset 
depends on the Ericson parameter as ${\mathcal N}=\left[{\mathcal E}\right]+1$, where $\left[\cdots\right]$ indicates integer part. 
This reduced subspace is defined by a projection operator $P_{n_0}^{({\mathcal N})}=\sum_{n=1}^{\mathcal N} \left|\Psi_{n_0+n}^{(0)}
\right\rangle\left\langle\Psi_{n_0+n}^{(0)}\right|$ where $n_0=1,\cdots N-{\mathcal N}$.

Next, we project Eq. (\ref{secular}) in the $P_{n_0}^{({\mathcal N})}$ subspace. The corresponding matrix elements of the reduced 
effective Hamiltonian $H_{eff,n_0}^{({\mathcal N})}(-k,\gamma)=P_{n_0}^{({\mathcal N})}H_{eff}(-k,\gamma)P_{n_0}^{({\mathcal N})}$ 
are expressed in terms of the eigenvalues $\{E_n^{(0)}\}$ and eigenvectors $\{\left|\Psi_n^{(0)}\right\rangle\}$ of the isolated system
which belong to the $P_{n_0}^{({\mathcal N})}$ subspace, i.e.
\begin{align}
\label{Heff_red}
\left[H_{eff,n_0}^{\left(\mathcal{N}\right)}\left(-k,\gamma\right)\right]_{nl}= & E_{n}^{\left(0\right)}\delta_{nl}
-\imath\gamma\sum_{d}\Braket{\Psi_{n}^{\left(0\right)}|e_{d}}\Braket{e_{d}|\Psi_{l}^{\left(0\right)}}\nonumber\\
+&\frac{w^{2}e^{-\imath k}}{t_{L}}\sum_{m}\Braket{\Psi_{n}^{\left(0\right)}|e_{m}}\Braket{e_{m}|\Psi_{l}^{\left(0\right)}}
\end{align}
where the indexes $d,m$ run over the position of the target(s) and the leads respectively. The potential WETAC pairs $(E_{WETAC},
\gamma_{WETAC})$ are associated with the real roots of the reduced secular equation $\zeta_{n_0}^{({\mathcal N})}(E,\gamma)
\equiv\det\left(H_{eff,n_0}^{\left(\mathcal{N}\right)}(-k,\gamma)-E(k)\right)=0$. Below, whenever not explicitly indicated, we shall 
assume that the analysis applies for all subspaces $n_0$.

Out of all possible pairs $(E_{WETAC},\gamma_{WETAC})$ which are solutions of the secular equation $\zeta^{({\mathcal N})}
(E,\gamma)=0$ we consider only the ones that satisfy the following ``proximity'' constrains: (a) $E_{WETAC}\in [E_{min}^{(0)}-
\delta, E_{max}^{(0)}+\delta]$ where $E_{min/max}^{(0)}$ are the borders of the energy interval associated with the eigenmodes 
of the reduced subspace $P^{({\mathcal N})}$ and $\delta\ll\Delta$; and (b) the evaluated $\gamma_{WETAC}\in {\bar \gamma}
[1-\eta,1+\eta]$ where ${\bar \gamma}$ is the loss strength (conductivity) of the target and $\eta$ is a tolerance level of our 
knowledge of its loss-strength. The corresponding subspace $P_{\rm WETAC}^{({\mathcal N})}$ which lead to a secular equation 
with solutions $(E_{WETAC},\gamma_{WETAC})$ that satisfy the above two constrains constitute a good basis for the description 
of WETACs. The WETAC waveforms $\left|I_{\rm WETAC}\right\rangle$ correspond to the eigenvector $\left|\alpha^{({\mathcal 
N})}_{\rm max}\right\rangle$ associated with the maximum eigenvalue $\alpha^{({\mathcal N})}_{\rm max}=1$ of the projected 
absorption operator ${\mathcal A}^{({\mathcal N})}$. The latter is given by Eq. (\ref{A_matrix}) with $G$ substituted by 
$G^{({\mathcal N})}\equiv P_{\rm WETAC}^{({\mathcal N})}GP_{\rm WETAC}^{({\mathcal N})}$.

{\it Ergodic WETACs --}In many practical situations, it is impossible to have information about the eigenmode amplitudes at 
the position of the target(s). We have therefore relax further the WETAC scheme by substituting in Eq. (\ref{Heff_red}) for 
$H_{eff}^{({\mathcal N})}$ (and consequently in $G^{({\mathcal N})}$), the eigenmode amplitudes at the position of the target(s) 
with their ergodic limit i.e. $\Braket{e_{d}|\Psi_{l}^{\left(0\right)}}\sim 1/\sqrt{N}$. This approximation is justified for chaotic cavities 
where typically the modes are ergodically distributed over the enclosure. We shall refer to this algorithm as the {\it ergodic} WETAC.

Below we test the proposed schemes for cavities with ${\mathcal E}\ll 1$ and ${\mathcal E}>1$ as well as for cavities with and 
without TR-invariance. 

{\it Isolated Resonances--} When ${\mathcal E}< 1$ the effective dimensionality of the projected subspaces is ${\mathcal N}=1$ 
and thus $P_n^{({\mathcal N}=1)}=\left|\Psi_n^{\left(0\right)}\right\rangle \left\langle\Psi_n^{\left(0\right)}\right|$ for $n=1,\cdots, N$. 
It turns out that in this case the evaluation of $H_{eff}^{({\mathcal N}=1)}$ requires only the knowledge of the field {\it intensities} 
at the position of the targets and at the position of the antennas, see Eq. (\ref{Heff_red}). A potential pair $(E_{WETAC}, 
\gamma_{WETAC})$ is calculated from the reduced secular equation $\zeta^{({\mathcal N})}(E,\gamma)=\left({\mathcal R}e\left[
\zeta^{({\mathcal N})}\right], {\mathcal I}m\left[\zeta^{({\mathcal N})}\right]\right)=(0,0)$. The pair is accepted as a WETAC solution 
if it satisfies the proximity constraints mentioned above. In this case the subspace $P_n^{({\mathcal N}=1)}$ is identified as 
$P_{WETAC}^{({\mathcal N}=1)}$ and is used for the evaluation of the WETAC field via ${\mathcal A}^{({\mathcal N}=1)}$. We get 
(see supplement \cite{supplement})
\begin{equation}
\label{I_WETAC_weak} 
\left|\mathcal{I}_{WETAC}\right\rangle \propto W^{T}\left|\Psi^{\left(0\right)}\right\rangle
\end{equation} 
where we have used  Eq. (\ref{A_matrix}) with the substitution of $G$ with $G^{({\mathcal N}=1)}=\left[\left\langle \Psi^{\left(0\right)}
\right|G\left|\Psi^{\left(0\right)}\right\rangle \right]\left|\Psi^{\left(0\right)}\right\rangle \left\langle \Psi^{\left(0\right)}\right|$.

At Figs. \ref{fig2}a,b,c we show the numerical results for ${\mathcal P}(\alpha)$ (see staircase blue line) for $N_d=1,2,4$ respectively. 
These distributions have been generated over a GOE ensemble of $H_0$ (for a fixed loss-strength ${\bar\gamma}$) by substituting 
Eq. (\ref{I_WETAC_weak},) together with the value of $E_{WETAC}$ satisfying the proximity constraints, in Eq. (\ref{absorbance}) for 
the numerical evaluation of the absorbance. 

\begin{figure}[h]
\includegraphics[width=1\columnwidth,keepaspectratio,clip]{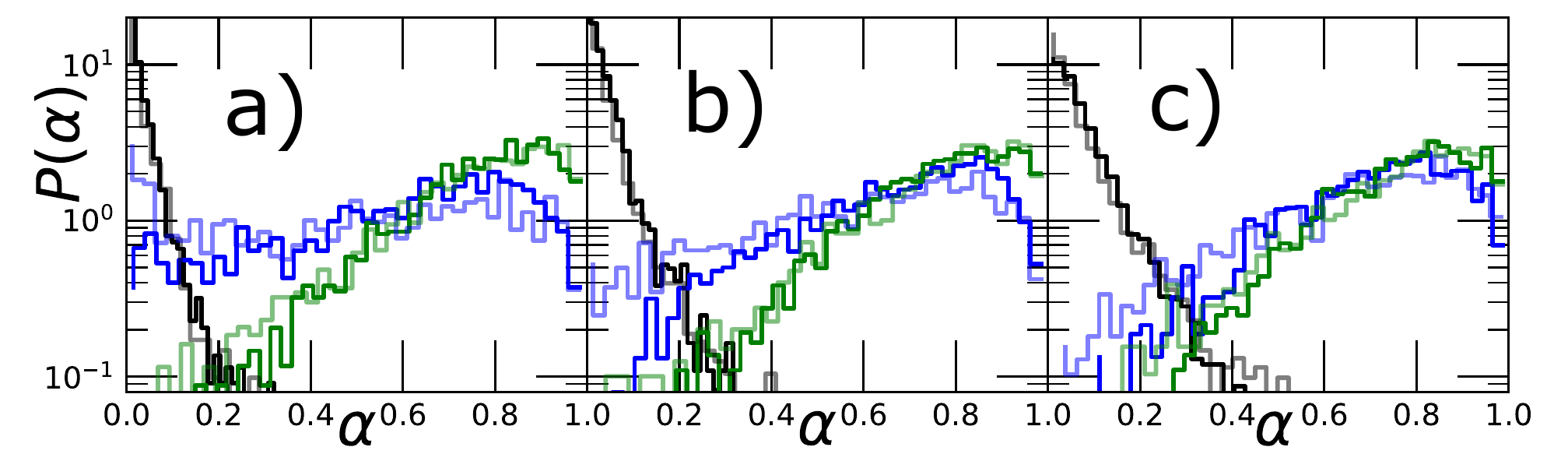}
\caption{ (color online) Probabilities of absorbance ${\mathcal P}(\alpha)$ for a GOE/GUE cavities (dark/light lines) 
with $w=-1$ and ${\mathcal E}\approx1.4/{\mathcal E}\approx1.7$. (a) One lossy target; (b) two lossy targets; and 
(c) four lossy targets. The loss tolerance in all cases is $\eta=0.3$. All other parameters are the same as in Fig. \ref{fig2}.
}
\label{fig3}
\end{figure}

We proceed with the theoretical evaluation of ${\mathcal P}(\alpha)$. Using Eqs.~(\ref{A_matrix},\ref{absorbance}) in the 
$P_{WETAC}^{({\mathcal N}=1)}$-space, we get the following expression for the absorbance (see supplement \cite{supplement})
\begin{align}
\alpha\left(k_{WETAC},\bar{\gamma}\right)= & \frac{4\bar{\gamma}\gamma_{WETAC}}{\left(\bar{\gamma}+
\gamma_{WETAC}\right)^{2}}.
\label{absorption_strength}
\end{align}
For simplicity, we assume that $\gamma_{WETAC}$ is a gaussian random variable with the same mean and variance with 
the one associated with a box distribution ${\bar \gamma} [1-\eta, 1+\eta]$. It is then straightforward to show that
\begin{align}
\mathcal{P}\left(\alpha\right)\propto & \sum_{\sigma=\pm}\frac{e^{-\frac{3}{2}\left(1-\alpha\right)\left(\frac{\alpha_{-\sigma}}{\alpha\eta}\right)^{2}}}{\sqrt{1-\alpha}\left(\alpha_{\sigma}-\alpha\right)}
\label{full_knowledge}
\end{align}
where $\alpha_{\sigma}\equiv2\left(1+\sigma\sqrt{1-\alpha}\right)$. The theoretical prediction Eq. (\ref{full_knowledge}) is plotted in 
Figs. \ref{fig2}a,b,c with a blue solid lines. 

In Fig. \ref{fig2} we also show the distribution of absorbances associated with incident waveforms $\left|\mathcal{I}_R\right\rangle=
\left|\alpha_{max}(k)\right\rangle$ corresponding to eigenvectors of ${\mathcal A}(k,{\bar \gamma})$ associated with the maximum 
eigenvalue, and random wavevector taken from a box distribution in the interval $k\in[0,\pi]$. We find a fast decay of ${\mathcal P}
(\alpha_{\rm max})$ for large $\alpha$ indicating that majority of these waveforms will miss the lossy target, see staircase black line. 
Notice that any other random waveform will be much less efficient. In the same figure we also plot the theoretical results for 
${\mathcal P}(\alpha_{\rm max})$, see continuous black lines \cite{supplement}.

Let us now, analyze the efficiency of the ergodic WETAC scheme. An optimal impedance match condition implies that the WETAC pair 
$(E_{WETAC},\gamma_{WETAC})$ must satisfy a flux-balance relation $\gamma_{WETAC}\sum_{d}\left|\left\langle e_{d}\right.\left|
\Psi_n^{\left(0\right)}\right\rangle \right|^{2}\sim v_g \sum_{m}\left|\left\langle e_{m}\right.\left|\Psi_n^{\left(0\right)}\right\rangle \right|^{2}$ 
where $v_g$ is the group velocity at $E=E_{WETAC}$. This relation allows us to write $\gamma_{WETAC}={\tilde \gamma}_{WETAC} 
\frac{N_d}{Y}$ where $Y\equiv N\sum_{d}\left|\left\langle e_{d}\right.\left|\Psi_n^{\left(0\right)}\right\rangle \right|^{2}$ and ${\tilde 
\gamma}_{WETAC} =\gamma_{WETAC}\left|_{\left|\langle e_{d}|\Psi^{\left(0\right)}\rangle\right| \approx1/\sqrt{N}}\right.$. We interpret 
${\tilde \gamma}_{WETAC}$ as the WETAC value for the loss-strength under the ergodic hypothesis for the field intensities at the position 
of the target(s) i.e. $\left|\langle e_{d}|\Psi^{\left(0\right)}\rangle\right| \approx1/\sqrt{N}$. Then Eq. (\ref{absorption_strength}) for the 
absorbance is rewritten as
\begin{align}
\alpha\left(k_{CPA},\bar{\gamma}\right)= & \frac{4\bar{\gamma}\tilde{\gamma}_{WETAC}\frac{N_{d}}{Y}}{\left(\bar{\gamma}+
\tilde{\gamma}_{WETAC}\frac{N_{d}}{Y}\right)^{2}}
\label{modified_absorption}
\end{align} 
where now ${\tilde \gamma}_{WETAC}\in {\bar \gamma} [1-\eta,1+\eta]$. In the large-$N$ limit, $Y$ satisfies the generalized Porter-Thomas 
distribution $P_{Y}\left(y\right)\propto e^{-y/2}y^{N_{d}/2-1}$ while we can further assume that $\tilde{\gamma}_{WETAC}$ is an independent 
random variable that obeys a Gaussian distribution. Using Eq.~(\ref{modified_absorption}) we get the distribution of the absorbances
\begin{align}
\tilde{P}\left(\alpha\right)\propto & \sum_{j=1,2}\frac{\alpha_{j}\left(1+\alpha_{j}\right)}{\alpha\left(1-\alpha_{j}\right)}I_{j}\left(N_d,\eta\right),
\label{ergodic}
\end{align}
where $I_{j}\left(N_d,\eta\right)\equiv\int_{0}^{\infty}\mathrm{d}x\exp\left[-\frac{3}{2\eta^{2}}\left(\frac{\alpha_{j}}{N_d}x-1\right)^{2}-\frac{x}{2}
\right]x^{N_d/2}$
and $\alpha_{j}\equiv\left(2-\alpha+\left(-1\right)^{j}2\sqrt{1-\alpha}\right)/\alpha$. In Fig. \ref{fig2} we plot with blue solid line the theoretical result 
Eq. (\ref{ergodic}) together with the numerical calculations (blue staircase line) for ergodic WETAC scheme. In comparison with the actual 
WETAC, the efficiency of the {\it ergodic WETAC} scheme to deliver the energy of the incident waveform at the lossy target is (obviously)
reduced. The ergodic WETAC scheme is, nevertheless, far superior to the random incident waves (black lines). The level of efficiency is 
improved further when more lossy targets $N_d>1$ are included in the complex enclosure, see Fig. \ref{fig2}b,c. The improvement is a direct
consequence of the validation of the ergodic hypothesis in the limit of many targets $1\ll N_d\ll N$.

{\it Overlapping resonances --} In this case the projected space is enlarged i.e. ${\mathcal N}>1$. Below we consider the example case 
of $M=2$ and ${\mathcal E}=1.4$ corresponding to ${\mathcal N}=2$. The projection operator takes the form $P^{({\mathcal N}=2)}=
\left|\Psi_n^{(0)}\right \rangle\langle \left. \Psi_{n}^{(0)}\right|+\left|\Psi_{n+1}^{(0)}\right \rangle\langle \left. \Psi_{n+1}^{(0)}\right|$ for all 
subsequent modes $E_n^{(0)},E_{n+1}^{(0)}$ of the isolated cavity $H_{0}$. The potential WETAC pairs $\left(k_{WETAC},\gamma_{WETAC}
\right)$ are obdained via Eqs. (\ref{secular},\ref{Heff_red}). Furthermore, the implementation of the proximity conditions 
allow us to single out the actual WETAC pairs and the corresponding WETAC subspaces ${\mathcal P}_{WETAC}^{(2)}$. 

The design of the WETAC waveforms requires the diagonalization of the reduced absorption matrix ${\mathcal A}^{({\mathcal N}=2)}$ in 
the WETAC subspaces ${\mathcal P}_{WETAC}^{(2)}$. The latter can be calculated using Eqs. (\ref{A_matrix}, \ref{Heff_red}). The eigenvector 
$\left|\alpha_{\rm max}\right \rangle$ associated with the eigenvalue $\alpha=1$ give us the desired WETAC $\left| {\mathcal I}_{\rm WETAC}
\right\rangle=\left|\alpha_{\rm max}\right\rangle$. For the case of one lossy target at position $d_0$ one has $\left|\mathcal{I}_{WETAC}
\right\rangle \propto\left|u_{d_{0}}\right\rangle $ (see Eq.~(\ref{A_matrix})), which in the ${\mathcal P}_{WETAC}^{(2)}$ space reads (see
supplement \cite{supplement})
\begin{align}
\left|\mathcal{I}_{WETAC}\right\rangle \propto & \left(\left[H_{eff}^{({\mathcal N}=2)}\right]_{2,2}-E_{WETAC}\right)
W^{T}\left|\Psi_{1}^{\left(0\right)}\right\rangle \nonumber\\
-& \left[H_{eff}^{({\mathcal N}=2)}\right]_{2,1}W^T\left|\Psi_{2}^{\left(0\right)}\right\rangle.
\end{align}
where $H_{eff}^{({\mathcal N}=2)}$ has been evaluated at $(k_{WETAC}, \gamma_{WETAC})$, see Eq. (\ref{Heff_red}). When $N_d>1$ 
the incident waveforms $\left|\mathcal{I}_{WETAC}\right\rangle$ are evaluated numerically using the aforementioned WETAC algorithm. 
In Figs. \ref{fig3}a,b,c we report our numerical results for the distribution of absorbances ${\mathcal P}(\alpha)$ when a WETAC incident 
wave is launched towards the complex cavity (green staircase) with $N_d=1,2,4$ lossy targets, respectively. At the same subfigures 
we report also the ${\mathcal P}(\alpha)$ associated with an ergodic WETAC (blue staircase). The two approaches converge rapidly to the 
same distribution as $N_d$ increases. As a reference we also show the distribution ${\mathcal P}(\alpha)$ for the case of $\left|{\mathcal 
I}_R\right\rangle$ incident waveforms (black staircase).

{\it WETACs for cavities with broken TR-invariance --} In Fig. \ref{fig3} we also report ${\mathcal P}(\alpha)$ for enclosures with broken
TR-symmetry. The corresponding incident waveforms have been generated using the same WETAC scheme as above for ${\mathcal E}=
1.7$. We find that the WETAC (light green staircase) and the ergodic WETAC (light blue staircase) schemes demonstrate the same level of 
efficiency as in the GOE case. Light black staircase lines indicate the ${\mathcal P}(\alpha)$ generated from an ensemble of $\left|{\mathcal 
I}_R\right\rangle$ incident waveforms and it is shown for comparison.

{\it Conclusions --} We have proposed a statistical algorithm that allow us to design waveforms that deliver, with high probability, large 
portion of their energy in weakly lossy targets which are embedded inside chaotic enclosures. There are many open questions that need 
further investigation. For example, can we guarantee simultaneous multiple strikes? What is the effect of a weakly lossy background? 
How non-universal features (like scars) can be utilized for better performance? These questions will be the theme of future research in 
WETAC shaping.

{\it Acknowledgments --} We acknowledge influential discussions with Dr. A. Nachman who shaped the direction of this research activity. We  
thank Profs. S. Anlage, H. Cao, H. Schanz and B. Shapiro for useful discussions on WETAC design. We also thank Prof. Y. Fyodorov for
pointing to us the reference [S1] which allowed us to derive Eq. (S12).


\vspace*{2cm}
\pagebreak
\widetext
\begin{center}
\textbf{\large Supplemental Materials}
\end{center}

\setcounter{equation}{0}
\setcounter{figure}{0}
\setcounter{table}{0}
\setcounter{page}{1}
\makeatletter
\renewcommand{\theequation}{S\arabic{equation}}
\renewcommand{\thefigure}{S\arabic{figure}}
\renewcommand{\bibnumfmt}[1]{[S#1]}
\renewcommand{\citenumfont}[1]{S#1}

\section{Derivation of Eq. (4)}

First we rewrite Eq.~(2) in the main text for the scattering matrix
as
\begin{align}
S= & -\hat{1}+2W^{T}\frac{1}{A_{H}+\Lambda+WW^{T}}W,\label{eq: S_new}
\end{align}
where $A_{H}=-A_{H}^{\dagger}=\frac{1}{\imath\sin k}\left[t_{L}\left(H_{0}-E\right)+\cos k\cdot WW^{T}\right]$
and $\Lambda=\Lambda^{\dagger}=-\frac{t_{L}}{\sin k}\Gamma_{0}$ with
$\Gamma_{0}=\gamma\sum_{d}\left|e_{d}\right\rangle \left\langle e_{d}\right|$.

Using Eq.~(\ref{eq: S_new}), the absorption operator $\mathcal{A}=1-S^{\dagger}S$
can be obtained as
\begin{align}
\mathcal{A}= & 2W^{T}\frac{1}{A_{H}+\Lambda+WW^{T}}W+2W^{T}\frac{1}{-A_{H}+\Lambda+WW^{T}}W\nonumber \\
 & -4W^{T}\frac{1}{-A_{H}+\Lambda+WW^{T}}WW^{T}\frac{1}{A_{H}+\Lambda+WW^{T}}W.\label{eq: A_intem}
\end{align}
We also have the identity 
\begin{align}
\frac{1}{A_{H}+\Lambda+WW^{T}}= & \frac{1}{-A_{H}+\Lambda+WW^{T}}\nonumber \\
+ & \frac{1}{-A_{H}+\Lambda+WW^{T}}\left(-2A_{H}\right)\frac{1}{A_{H}+\Lambda+WW^{T}}\label{eq: identity}
\end{align}
Substituting Eq.~(\ref{eq: identity}) into Eq.~(\ref{eq: A_intem}),
we obtain
\begin{align}
\mathcal{A}= & 4W^{T}\frac{1}{-A_{H}+\Lambda+WW^{T}}\Lambda\frac{1}{A_{H}+\Lambda+WW^{T}}W,
\end{align}
or equivalently

\begin{align}
\mathcal{A}= & -4\gamma\frac{\sin k}{t_{L}}\sum_{d}\left|u_{d}\right\rangle \left\langle u_{d}\right|,\:\left|u_{d}\right\rangle =W^{T}G\left|e_{d}\right\rangle 
\end{align}
where $G=\left[H_{eff}^{\dagger}-E\right]^{-1}$ with $H_{eff}=H_{0}-\imath\Gamma_{0}+\frac{e^{\imath k}}{t_{L}}WW^{T}$

\section{Derivation of Eq.(8) and Eq. (9)}

In the case of good cavities, using Eqs. (6) and (7) the WETAC pair
$\left(E_{WETAC},\gamma_{WETAC}\right)$ can be given explicitly as 

\begin{align}
\frac{E_{WETAC}-E^{\left(0\right)}}{E_{WETAC}}= & \frac{1}{2}\left(\frac{w}{t_{L}}\right)^{2}\Sigma^{m}\nonumber \\
\frac{\gamma_{WETAC}}{v_{g}\left(k_{WETAC}\right)}= & \frac{1}{2}\left(\frac{w}{t_{L}}\right)^{2}\frac{\Sigma^{m}}{\Sigma^{d}}\label{eq: k_good}
\end{align}
where $\Sigma^{m}\equiv\sum_{m}\left|\left\langle e_{m}\right.\left|\Psi^{\left(0\right)}\right\rangle \right|^{2}$,
$\Sigma^{d}\equiv\sum_{d}\left|\left\langle e_{d}\right.\left|\Psi^{\left(0\right)}\right\rangle \right|^{2}$
and the pair $\left(E^{\left(0\right)},\left|\Psi^{\left(0\right)}\right\rangle \right)$
is defined as $H_{0}\left|\Psi^{\left(0\right)}\right\rangle =E^{\left(0\right)}\left|\Psi^{\left(0\right)}\right\rangle $
with the normalized $\left|\Psi^{\left(0\right)}\right\rangle $ being
in the space $P_{WETAC}^{\left(\mathcal{N}=1\right)}$.

Under the single-mode approximation, the absorption operator in Eq.
(4) is given as

\begin{align}
\mathcal{A}^{\left(\mathcal{N}=1\right)}\left(k_{WETAC},\bar{\gamma}\right)= & -4\bar{\gamma}\frac{\sin k_{WETAC}}{t_{L}}\Sigma^{d}\left|G_{0}\right|^{2}\left[W^{T}\left|\Psi^{\left(0\right)}\right\rangle \left\langle \Psi^{\left(0\right)}\right|W\right],\label{eq: A_good}
\end{align}
where $G_{0}\equiv\left\langle \Psi^{\left(0\right)}\right|G\left(k_{WETAC},\bar{\gamma}\right)\left|\Psi^{\left(0\right)}\right\rangle $
can be further approximated to be 
\begin{align}
G_{0}\approx & \left[\imath\bar{\gamma}\Sigma^{d}+\frac{e^{-\imath k_{WETAC}}}{t_{L}}w^{2}\Sigma^{m}-\left(E_{WETAC}-E^{\left(0\right)}\right)\right]^{-1}.\label{eq: G0}
\end{align}
Eq.~(\ref{eq: A_good}) implies the WETAC field 
\begin{align}
\left|\mathcal{I}_{WETAC}\right\rangle \propto & W^{T}\left|\Psi^{\left(0\right)}\right\rangle 
\end{align}
$i.e.,$ Eq. (8) of the main text. Correspondingly, the absorbance
$\alpha\left(k_{WETAC},\bar{\gamma}\right)\approx\frac{\left\langle \mathcal{I}_{WETAC}\right|\mathcal{A}^{\left(\mathcal{N}=1\right)}\left(k_{WETAC},\bar{\gamma}\right)\left|\mathcal{I}_{WETAC}\right\rangle }{\left\langle \mathcal{I}_{WETAC}\right.\left|\mathcal{I}_{WETAC}\right\rangle }$
is given as 
\begin{align}
\alpha\left(k_{WETAC},\bar{\gamma}\right)= & \frac{4\bar{\gamma}\gamma_{WETAC}}{\left(\bar{\gamma}+\gamma_{WETAC}\right)^{2}}
\end{align}
which is Eq. (9) of the main text. In the derivation we have used
Eqs.~(\ref{eq: k_good}) and (\ref{eq: G0}).

\section{Absorbances for $\left|\mathcal{I}_{R}\right\rangle $}

We present here a theoretical analysis of the absorbance probability
distribution $\mathcal{P}(\alpha)$ for the case when the incident
waveforms are given by an ensemble of $\left|\mathcal{I}_{R}\right\rangle $
(see the main text). We restrict the discussion to the case of one
loss target at site $d_{0}$. From Eq.~(\ref{A_matrix}), we see
that there is only one nonzero eigenvalue of the absorption matrix
${\mathcal{A}}$ which is given as 
\begin{equation}
\alpha=-4\gamma\frac{\sin k}{t_{L}}\left\langle u_{d_{0}}\right|\left.u_{d_{0}}\right\rangle \approx-4\bar{\gamma}\frac{w^{2}}{t_{L}}\sum_{m}\left[H_{0}^{-1}\right]_{md_{0}}^{2}\label{randomalpha}
\end{equation}
where we have assumed that the incident energy is in the middle of
the band and the summation is over the number of leads $m=1,\cdots,M$.

We assume further that the \textit{off-diagonal} entries $\left[H_{0}^{-1}\right]_{md_{0}}$
are statistically independent. Thus the study of the distribution
$\mathcal{P}(\alpha)$ collapses to the study of the distribution
of the random variables $x=\left[H_{0}^{-1}\right]_{md_{0}}$, ($H_{0}$
is a GOE random matrix, $i.e.,$ $H_{0}$ is a random $N\times N$
real symmetry matrix satisfying the distribution $\mathcal{P}\left(H_{0}\right)\propto\exp\left(-\frac{N}{4}\mathrm{Tr}H_{0}^{2}\right)$
). In the large matrix-size limit $N\rightarrow\infty$, it has been
shown that $\mathcal{P}\left(x\right)=\frac{2}{\pi^{2}\left(1+x^{2}\right)}\left[1+\frac{\mathrm{arsinh\left(x\right)}}{x\sqrt{1+x^{2}}}\right]$\cite{Fyodorov_Nock}.
Notice that when $x\rightarrow\pm\infty$, $\mathcal{P}(x)\propto1/x^{2}$.
Using this information in Eq. (\ref{randomalpha}) we eventually get
\begin{align}
\hat{\mathcal{P}}\left(\alpha\right)\propto & \int_{-\infty}^{\infty}\mathrm{d}xT\left(x\right)^{M}\exp\left(-\imath x\frac{t_{L}\alpha}{4\bar{\gamma}w^{2}}\right)
\end{align}
where $T\left(x\right)=\int_{-\infty}^{\infty}dy\frac{1}{1+y^{2}}\left[1+\frac{\mathrm{arsinh\left(y\right)}}{y\sqrt{1+y^{2}}}\right]\exp\left(-\imath xy^{2}\right)$.

\section{Derivation of Eq.(13) }

In the case of overlapping resonances, there exists two eigenvectors
$\left|\Psi_{1}^{\left(0\right)}\right\rangle $ and $\left|\Psi_{2}^{\left(0\right)}\right\rangle $
of the Hamiltonian $H_{0}$ within one $\mathcal{P}_{WETAC}^{\left(2\right)}$
space. The WETAC pair $\left(E_{WETAC},\gamma_{WETAC}\right)$ is
determined through $\zeta^{\left(2\right)}\left(E_{WETAC},\gamma_{WETAC}\right)=0$,
or explicitly 
\begin{align}
\det U= & 0,\:U\equiv\left(\begin{array}{cc}
\left[H_{eff}^{\left(\mathcal{N}=2\right)}\right]_{11}-E_{WETAC} & \left[H_{eff}^{\left(\mathcal{N}=2\right)}\right]_{12}\\
\left[H_{eff}^{\left(\mathcal{N}=2\right)}\right]_{21} & \left[H_{eff}^{\left(\mathcal{N}=2\right)}\right]_{22}-E_{WETAC}
\end{array}\right)\label{eq: secular_N2}
\end{align}
where $\left[H_{eff}^{\left(\mathcal{N}=2\right)}\right]_{nl}$ is
given in Eq. (7) of the main text. We consider the case of one lossy
target at position $d_{0}$, when the absorption operator is given
as $\mathcal{A}\left(k,\gamma\right)=-4\gamma\frac{\sin k}{t_{L}}\left|u_{d_{0}}\right\rangle \left\langle u_{d_{0}}\right|$
with $\left|u_{d_{0}}\right\rangle =W^{T}G\left|e_{d_{0}}\right\rangle $.
In the $\mathcal{P}_{WETAC}^{\left(2\right)}$ space, the WETAC field
reads
\begin{align}
\left|\mathcal{I}_{WETAC}\right\rangle \propto & W^{T}G^{\left(\mathcal{N}=2\right)}\left|e_{d_{0}}\right\rangle 
\end{align}
or explicitly 
\begin{align}
\left|\mathcal{I}_{WETAC}\right\rangle \propto & W^{T}\left|\Psi_{1}^{\left(0\right)}\right\rangle g_{1}+W^{T}\left|\Psi_{2}^{\left(0\right)}\right\rangle g_{2}\label{eq: I_N2}
\end{align}
where $g_{j}=\sum_{n=1}^{2}G_{jn}^{\left(\mathcal{N}=2\right)}\left\langle \Psi_{n}^{\left(0\right)}\right|\left.e_{d_{0}}\right\rangle $
and $G_{jn}^{\left(\mathcal{N}=2\right)}\equiv\left\langle \Psi_{j}^{\left(0\right)}\right|G^{\left(\mathcal{N}=2\right)}\left|\Psi_{n}^{\left(0\right)}\right\rangle $.
In addition, the $2\times2$ matrix $\left(G_{jn}^{\left(\mathcal{N}=2\right)}\right)$
in the $\mathcal{P}_{WETAC}^{\left(2\right)}$ space can be given
as 
\begin{align}
\left(G_{jn}^{\left(\mathcal{N}=2\right)}\right)= & \left[U+2\imath\begin{pmatrix}\Gamma_{0}^{11} & \Gamma_{0}^{12}\\
\Gamma_{0}^{21} & \Gamma_{0}^{22}
\end{pmatrix}\right]^{-1}\label{eq: G_N2}
\end{align}
where $\Gamma_{0}^{jn}=\left\langle \Psi_{j}^{\left(0\right)}\right|\Gamma_{0}\left|\Psi_{n}^{\left(0\right)}\right\rangle $.
Using Eqs.~(\ref{eq: secular_N2}) and (\ref{eq: G_N2}), we can
easily obtain 
\begin{align}
\frac{g_{2}}{g_{1}}= & -\frac{\left[H_{eff}^{\left(\mathcal{N}=2\right)}\right]_{21}}{\left[H_{eff}^{\left(\mathcal{N}=2\right)}\right]_{22}-E_{WETAC}}.\label{eq: ratio_N2}
\end{align}
Therefore combining Eq.~(\ref{eq: I_N2}) and Eq.~(\ref{eq: ratio_N2})
together, we finally reach Eq. (13) of the main text.

\end{document}